\documentclass[twocolumn,aps,superscriptaddress,showpacs]{revtex4}

\usepackage{amssymb}
\usepackage{amsmath}
\usepackage{mathrsfs}
\usepackage{graphicx}
\usepackage{subfigure}
\usepackage[normalem]{ulem}
\usepackage[dvips]{color}

\begin{document}

\title{Speed of sound and polytropic index in the Polyakov-loop Nambu-Jona-Lasinio model}
\author{He Liu}
\email{liuhe@qut.edu.cn}
\affiliation{Science School, Qingdao University of Technology, Qingdao 266520, China}
\affiliation{The Research Center of Theoretical Physics, Qingdao University of Technology, Qingdao 266033, China}
\author{Yong-Hang Yang}
\affiliation{Science School, Qingdao University of Technology, Qingdao 266520, China}
\affiliation{The Research Center of Theoretical Physics, Qingdao University of Technology, Qingdao 266033, China}
\author{Chi Yuan}
\affiliation{School of Mechanical and Automotive Engineering, Qingdao University of Technology,
Qingdao 266520, China}
\author{Min Ju}
\email{jumin@upc.edu.cn}
\affiliation{School of Science, China University of Petroleum (East China), Qingdao 266580, China}
\author{Xu-Hao Wu}
\email{wuhaoysu@ysu.edu.cn}
\affiliation{School of Science, Yanshan University, Qinhuangdao 066004, China}
\author{Peng-Cheng Chu}
\email{kyois@126.com}
\affiliation{Science School, Qingdao University of Technology, Qingdao 266520, China}
\affiliation{The Research Center of Theoretical Physics, Qingdao University of Technology, Qingdao 266033, China}
\date{\today}
\begin{abstract}
We investigate the speed of sound and polytropic index of quantum chromodynamics (QCD) matter in the full phase diagram based on a 3-flavor Polyakov-looped Nambu-Jona-Lasinio (PNJL) model. The speed of sound and polytropic index in isothermal and adiabatic cases all have a dip structure at the low chemical potential side of the chiral phase transition boundary, and these quantities reach their global minimum values at the critical endpoint (CEP) but are not completely zero, where the values in adiabatic are lightly greater than those in isothermal. Different from the speed of sound, the polytropic index rapidly form a peak after crossing the chiral phase transition boundary, which is observed for the first time. Along the hypothetical chemical freeze-out lines, the speed of sound rapidly decreases near the CEP, followed by a small spinodal behavior, while the polytropic index, especially in isothermal, exhibits a more pronounced and nearly closed to zero dip structure as it approaches the CEP. 
\end{abstract}


\maketitle
\section{Introduction}
\label{INTRO}
Exploring the phase structure of quantum chromodynamics (QCD) matter and searching for the signals of the phase transition are significant goals in both theoretical research and heavy-ion collision experiments. Lattice QCD (LQCD) simulations indicate the transition between the quark-gluon plasma (QGP) and the hadronic matter is a smooth crossover at nearly zero baryon chemical potential ($\mu_B$)~\cite{Aok06,Gup11,Bor14}. A first-order phase transition, with a critical endpoint (CEP) connecting with the crossover transformation, is predicted at large baryon chemical potential according to various effective models, such as the Nambu-Jona-Lasinio (NJL) model and the quark-meson (QM) model, as well as through advanced functional methods including the Dyson-Schwinger equation (DSE) and the functional renormalization group (FRG)~\cite{Sch99,Zhu00,Fu08,Fuk08,Qin11,Sch12,Che16,Liu16,Fu20,Liu21}. In order to find the signal of the QCD CEP at finite $\mu_B$, the Beam Energy Scan (BES) program is currently ongoing at the Relativistic Heavy-Ion Collider (RHIC)~\cite{Agg10,Ada142}. In the BES program, the STAR experiment has measured the energy dependence of observables that are sensitive to the CEP and/or first-order phase transition, including net-proton fluctuations~\cite{Ada21L,Abd21}, pion HBT radii~\cite{Ada15,Ada21C}, baryon directed flow~\cite{Ada141,Ada18}, intermittency of charged hadrons~\cite{Abd23B},and light nuclei yield ratio ($N_t\times N_p/N_d^2$)~\cite{Abd23L,Sun18,Sun21A,Sun21B}. Nonmonotonic energy dependencies were observed in all of these observables, and the energy ranges where peak or dip structures appear are around $\sqrt{sNN} \approx 7.7-39$ GeV. Those provoking observations are of great interest, and a more accurate measurement on BES-II in the near future will provide us more information about the QCD phase diagram.

The speed of sound ($c_s$), as one of fundamental properties of substance, can also convey QCD phase structure information. Lattice QCD, for instance, shows that the minimum value of the squared speed of sound ($c_s^2$) occurs in the crossover region with $T_0 = 154 \pm 9$ MeV at vanishing baryon chemical potential~\cite{Baz14}. The numerical results from the effective models~\cite{Mot20,He22,He23} suggest that the speed of sound is the global minimum at CEP, but it does not completely vanish in the mean field approximation. The speed of sound is one of the crucial physical quantities in hydrodynamics, which carries important information in describing the evolution of strongly interacting matter and final observables in heavy-ion collision experiments. The studies in~\cite{Cam11,Gar20,Sah20} indicate that the speed of sound as a function of charged particle multiplicity $\langle dN_{ch}/d\eta \rangle$ can be extracted from heavy-ion collision data. Recently, the authors~\cite{Sor21} estimate the value of $c_s$ as well as its logarithmic derivative with respect to the baryon number density and try to build a connection with the cumulants of the baryon number distribution in matter created in heavy-ion collisions to aid in detecting the QCD CEP. 

The first-order phase transition from the hadronic to quark matter at high baryon densities may also occur in the interior of massive neutron stars~\cite{Gle01,Web05}. Some recent studies have shown that quark-matter cores can appear in massive neutron stars~\cite{Ann20,Liu22,Liu237,Liu238}. Compared with the hadronic matter (HM), strange quark matter (SQM) is known to exhibit markedly different properties. For example, SQM at very high densities ($\rho_B \geq 40\rho_0$) is approximately scale invariant or conformal, whereas in HM the degree of freedom is smaller and the scale invariance is also violated by the breaking of chiral symmetry. These qualitative differences between HM and SQM can be reflected in the speed of sound, where $c_s$ takes the constant $c_s^2=1/3$ in the exactly conformal matter corresponding to SQM at high densities. However, $c_s^2$ in HM varies considerably: below saturation density, most hadronic models, such as chiral effective field theory, indicate $c_s^2 \ll 1/3$, while at higher densities the maximum of $c_s^2$ is predicted to be greater than 0.5~\cite{Gan09,Tew13}. Moreover, the behavior of $c_s$ as a function of baryon number density influences the mass-radius relation, the tidal deformability and gravitational wave, and thus is helpful in understanding the equation of state (EOS) of neutron star matter and QCD phase structure. A recent research in Ref.~\cite{Hua22} finds that the EOS with a pronounced peak in the speed of sound leaves a clear and unique signature in the main frequency of the postmerger gravitational wave (GW) spectrum, which can provide a sensitive probe of the hadron-quark phase transition in the dense core. 

On the other hand, the physical quantity that the polytropic index, defined as $\gamma \equiv \partial \ln P/\partial \ln \varepsilon$, can also convey the qualitative differences between HM and SQM. The polytropic index has the value $\gamma=1$ in conformal quark matter, while the hadronic models generically predict $\gamma=2.5$ around the saturation density~\cite{Kur10}. Therefore, some researches indicate the approximate rule that the polytropic index $\gamma<1.75$, taken as the average of its value at saturation density and the conformal limit, can be used as a good criterion for separating quark matter from hadronic matter in the massive neutron star core~\cite{Ann20,Liu22,Liu237,Liu238}. In Ref.~\cite{Han23}, authors suggest that a more "conservative" criterion $\gamma\leq 1.6$ and $c_s^2\leq0.7$ can be used to determine the possible onset of the exotic matter (likely made of quark matter). Recently, a new conformality criterion $d_c <0.2$, defined as being composed of $\gamma$ and $c_s$, is seen to be considerably more restrictive than the criterion $\gamma <1.75$ for quark matter in compact stars~\cite{Ann23}. Therefore, the polytropic index could provide a new probe of the QCD CEP or first-order phase transition in future experimental exploration.

Based on the significant findings mentioned above from recent years, in the present study, we focus on investigating the polytropic index of QCD matter in the full phase diagram based on a 3-flavor Polyakov-looped Nambu-Jona-Lasinio (PNJL) model. As a crucial intermediate quantity in the $\gamma$ calculation process and as a comparison to $\gamma$, we have also correspondingly calculated the full phase diagram properties of the speed of sound. The definitions of the speed of sound and polytropic index require specifying which properties of the system are considered constant. In this work, we mainly calculate the speed of sound and polytropic index in the adiabatic and isothermal cases and analyze the changing behavior of these quantities around the chiral phase transition boundary and CEP. By constructing the hypothetical chemical freeze-out lines, we can find that both the speed of sound and polytropic index are close to zero near the CEP. The difference is that the speed of sound rapidly decreases near the CEP, followed by a small spinodal behavior, while the polytropic index exhibits a more pronounced dip structure as it approaches the CEP.

\section{The theoretical model }
\label{MODEL}
The thermodynamic potential density of the 3-flavor PNJL model at finite temperature $T$ can be expressed as~\cite{Fuk08,Liu16}
\begin{eqnarray}\label{eq1}
\Omega_{\textrm{PNJL}} &=& \mathcal{U}(\Phi,\bar{\Phi},T)-2N_c\sum_{i=u,d,s}\int_0^\Lambda\frac{d^3p}{(2\pi)^3}E_i
\notag\\
&-&2T\sum_{i=u,d,s}\int\frac{d^3p}{(2\pi)^3}\{\ln[1+3\Phi e^{-\beta(E_i-\mu_i)}
\notag\\
&+&3\bar{\Phi}e^{-2\beta(E_i-\mu_i)}+e^{-3\beta(E_i-\mu_i)}]
\notag\\
&+&\ln[1+3\bar{\Phi} e^{-\beta(E_i+\mu_i)}
\notag\\
&+&3\Phi e^{-2\beta(E_i+\mu_i)}+e^{-3\beta(E_i+\mu_i)}]\}
\notag\\
&+&G_S(\sigma_u^2+\sigma_d^2+\sigma_s^2)-4K\sigma_u\sigma_d\sigma_s.
\end{eqnarray}
In the above, the temperature-dependent effective potential $\mathcal{U}(\Phi, \bar{\Phi}, T)$ as a function of the Polyakov loop $\Phi$ and $\bar{\Phi}$ is expressed as~\cite{Fuk08}
\begin{eqnarray}
\mathcal{U}(\Phi,\bar{\Phi},T) &=& -b \cdot T\{54e^{-a/T}\Phi\bar{\Phi}
+\ln[1-6\Phi\bar{\Phi}
\notag\\
&-&3(\Phi\bar{\Phi})^2+4(\Phi^3+\bar{\Phi}^3)]\},
\end{eqnarray}
with the parameters $a$ = 664 MeV and $b$ = 0.03$\Lambda^3$, which leads to simultaneous crossovers of chiral restoration and deconfinement around $T \approx$ 200 MeV. In Eq.\eqref{eq1}, the factor $2N_c$ with $N_c = 3$ represents the spin and color degeneracy, $\beta = 1/T$ represents the reciprocal of temperature, and $\mu_i$ denotes the chemical potential of quark with flavor $i$. In our calculation, we have set $\mu_u=\mu_d=\mu_s=\mu_B/3$, where $\mu_B$ is the baryon chemical potential. $G_S$ is the strength of the scalar coupling, and the $K$ term represents the six-point Kobayashi-Maskawa-t’Hooft (KMT) interaction which is required to break the axial $U(1)_A$ symmetry~\cite{Hoo76}. In the present calculation, we adopt the values of parameters given in Ref.~\cite{Hat94} as $G_S\Lambda^2$ = 3.67, $K\Lambda^5$ = 9.29, where $\Lambda$ = 631.4 MeV is the cutoff value in the momentum integration. The energy $E_i$ of quarks with flavor $i$ is expressed as $E_i(p)=\sqrt{p^2 +M_i^2}$, where $M_i$ represents the constituent quark mass. In the mean-field approximation (MFA), quarks can be taken as quasiparticles with constituent masses $M_i$, which is related to spontaneous chiral symmetry breaking. The constituent quark mass $M_i$ is determined by the gap equation of
\begin{eqnarray}
M_i &=& m_i-2G_S\sigma_i+2K\sigma_j\sigma_k,
\end{eqnarray}
where $m_i$ refers to the masses of the current quarks, which can be regarded as parameters. In our work, the values assigned are $m_u = 5.5$ MeV for the up quark, $m_d = 5.5$ MeV for the down quark and $m_s = 137.5$ MeV for the strange quark~\cite{Hat94}. The $\sigma_i=\langle\bar{q}_iq_i\rangle$ stands for quark condensate, which serves as the order parameter for the chiral phase transition. In a chirally symmetric environment, at low temperatures but high densities, quarks can also form diquark condensate $\langle q_iq_i\rangle$, breaking the color symmetry and leading to what is known as color superconductivity (CSC)~\cite{Rus05,Bub05,Gia05,Yua23}. At very high densities, quarks are believed to pair up in a unique manner that locks their color and flavor degrees of freedom together, resulting in a novel phase of QCD matter like the color-flavor locked (CFL) phase~\cite{Bai84,Alf01,Sho05}. The diquark condensate are indeed a vital component to consider in a comprehensive description of QCD matter under extreme conditions, especially concerning the finite density phase diagram. However, the color superconducting gap and critical temperature calculated using the perturbation theory are rather small~\cite{Bai84}, which is more likely applicable to the study of cold neutron stars. In our previous work~\cite{Chu23}, we have attempted to explore the role of CFL matter in quark stars. Simultaneously, the superconducting gap in strongly interacting matter, as calculated using the NJL model, was found to be significantly larger, with a value on the order of approximately 100 MeV~\cite{Sch99,Alf98,Rap98}. These studies emphasized the pivotal role of CSC in the QCD phase diagram and its potential significance in heavy-ion collisions~\cite{Iwa95,Pis99}. Thus, the future works could incorporate diquark condensates into the analysis to provide a deeper insight into the rich phase structure of QCD.

In the current work, we primarily focus on the quark chiral phase transition, and the values of $\sigma_u$, $\sigma_d$, $\sigma_s$, $\Phi$, $\bar{\Phi}$ in PNJL model can be calculated using the following equations
{\small
\begin{eqnarray}
\frac{\partial\Omega_{\textrm{PNJL}}}{\partial\sigma_u}
=\frac{\partial\Omega_{\textrm{PNJL}}}{\partial\sigma_d}
=\frac{\partial\Omega_{\textrm{PNJL}}}{\partial\sigma_s}
=\frac{\partial\Omega_{\textrm{PNJL}}}{\partial\Phi}
=\frac{\partial\Omega_{\textrm{PNJL}}}{\partial\bar{\Phi}}
=0.
\notag\\
\end{eqnarray}}

The pressure, number density, and entropy density can be derived using the thermodynamic relations in the grand canonical ensemble as
\begin{eqnarray}
P=-\Omega_{\textrm{PNJL}}, \quad \rho_B=-\frac{\partial\Omega_{\textrm{PNJL}}}{\partial\mu_B}, \quad s=-\frac{\partial\Omega_{\textrm{PNJL}}}{\partial T},
\end{eqnarray}
and energy density can be calculated as
\begin{eqnarray}
\varepsilon=-P+Ts+\mu_B\rho_B.
\end{eqnarray}

The speed of sound is the velocity of a longitudinal compression wave propagating through the medium, which is a fundamental property of substance~\cite{Sor21}. The general definition of speed of sound is $c_s^2=\partial P/\partial\varepsilon$. It is worth noting that calculating the speed of sound requires specifying which thermodynamic variables are kept constant. For the QGP created in relativistic heavy-ion collisions, it evolves with constant entropy per baryon $s/\rho_B$, so that using the adiabatic speed of sound $c_{s/\rho_B}$ is appropriate. Differently, the isothermal speed of sound $c_T$ is widely used in neutron star matter. However, the study in Ref.~\cite{Sor21} suggests that the cumulants of the baryon number distribution in heavy-ion collisions can be used to estimate the isothermal speed of sound squared and its logarithmic derivative with respect to the baryon number density. This result provides a new method for obtaining information about the QCD structure in heavy-ion collisions and neutron star studies. Using the Jacobian determinant method and thermodynamic relations, the $c_T^2$ and $c_{s/\rho_B}^2$ in terms of $T$ and $\mu_B$ are respectively written as
\begin{eqnarray}
c_T^2&=&\frac{\rho_B}{T(\frac{\partial s}{\partial \mu_B})_T+\mu_B(\frac{\partial \rho_B}{\partial \mu_B})_T},
\end{eqnarray}
and
{\small
\begin{eqnarray}
c_{s/\rho_B}^2&=&\frac{s\rho_B(\frac{\partial s}{\partial \mu_B})_T\!-s^2(\frac{\partial \rho_B}{\partial \mu_B})_T\!- \rho_B^2(\frac{\partial s}{\partial T})_{\mu_B}\!+\rho_B s(\frac{\partial \rho_B}{\partial T})_{\mu_B}}{(sT+\mu_B\rho_B)[(\frac{\partial s}{\partial \mu_B})_T(\frac{\partial \rho_B}{\partial T})_{\mu_B}\!-(\frac{\partial s}{\partial T})_{\mu_B}(\frac{\partial \rho_B}{\partial \mu_B})_T]}.
\notag\\
\end{eqnarray}
}

Meanwhile, the polytropic index is introduced in~\cite{Ann20} $\gamma \equiv \partial \ln P/\partial \ln \varepsilon $ as a criterion for separating hadronic matter from quark matter (or the exotic matter)~\cite{Ann20,Mal22,Han23,Liu237,Liu238}. For the isothermal and adiabatic processes, the polytropic index can be derived using the speed of sound formulas as
\begin{eqnarray}
\gamma_T=(\frac{\partial P}{\partial\varepsilon})_T/\frac{P}{\varepsilon}=\frac{\varepsilon}{P}c_T^2,
\end{eqnarray}
and
\begin{eqnarray}
\gamma_{s/\rho_B}=(\frac{\partial P}{\partial\varepsilon})_{s/\rho_B}/\frac{P}{\varepsilon}=\frac{\varepsilon}{P}c_{s/\rho_B}^2.
\end{eqnarray}
From the above two expressions, we can see that the speed of sound serves as an important intermediate quantity in the calculation of the polytropic index. The differences between these two quantities might lead to the polytropic index including more intricate features near the phase boundary, which could potentially provide a new probe for exploring the CEP.

\begin{figure*}[tbh]
\includegraphics[scale=0.45]{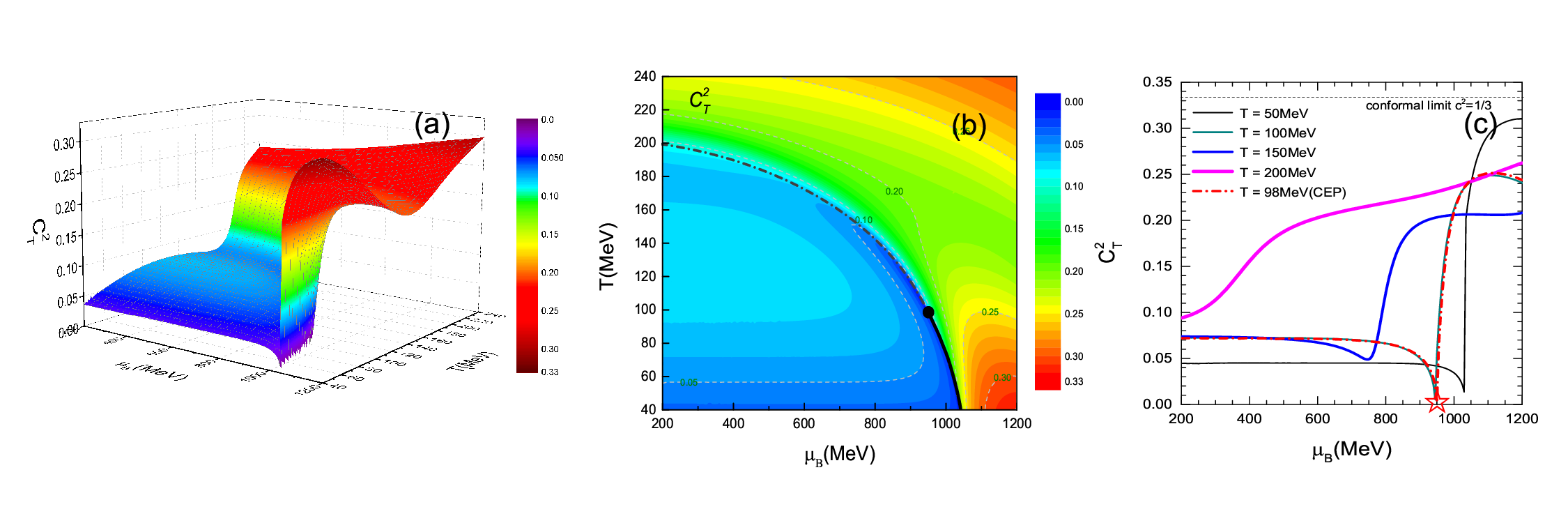}
\caption{The 3D plot (a) and contour map (b) of squared speed of sound $c_T^2$ in the $\mu_B-T$ plane based on a 3-flavor PNJL model, as well as the squared speed of sound $c_T^2$ as functions of $\mu_B$ (c) at $T = 50$, 100, 150, 200 MeV. In (b), the black dash-dotted and solid lines are respectively for the chiral crossover and first-order phase transition, the black dot connecting the chiral crossover and first-order phase transition represents the critical endpoint (CEP), and the gray dashed lines correspond to the contour of the squared speed of sound. In (c), the red star marks the baryon chemical potential of the CEP, and the black dashed line $c_s^2=1/3$ indicating the conformal matter limit is also shown for comparison.} \label{fig1}
\end{figure*}

\begin{figure*}[tbh]
\includegraphics[scale=0.45]{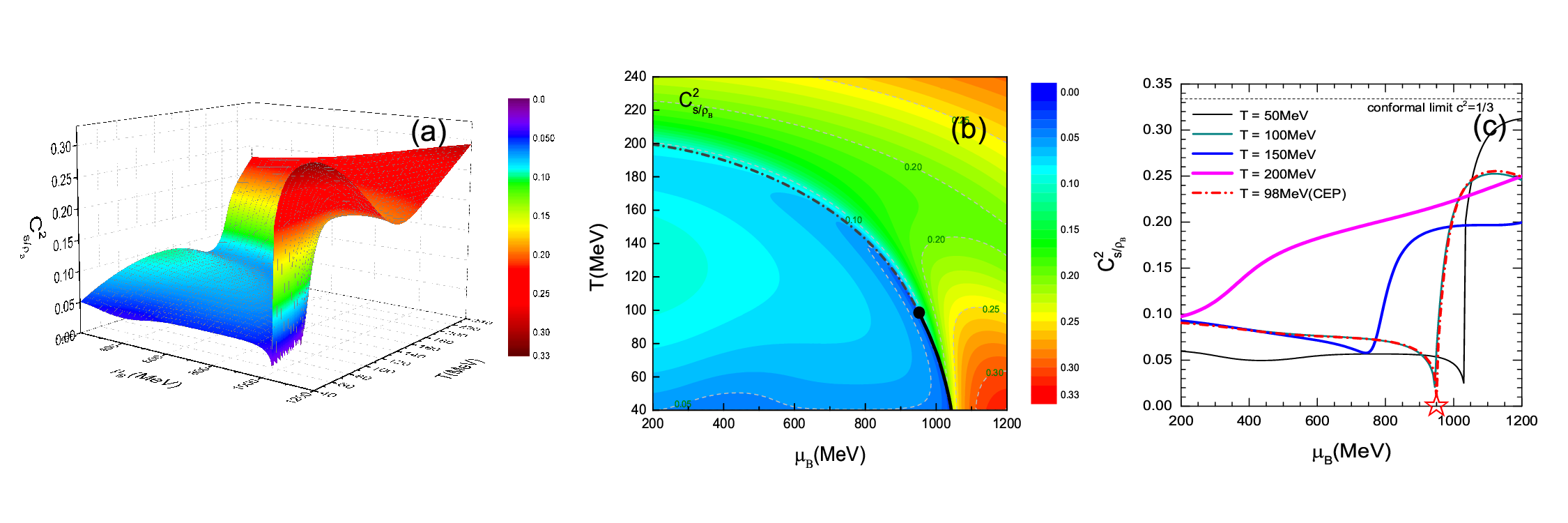}
\caption{The same as in Fig.~\ref{fig1}, but for the squared speed of sound $c_{s/\rho_B}^2$.}\label{fig2}
\label{fig2}
\end{figure*}

\section{Results and Discussions}
\label{RESULT}

\begin{figure*}[tbh]
\includegraphics[scale=0.45]{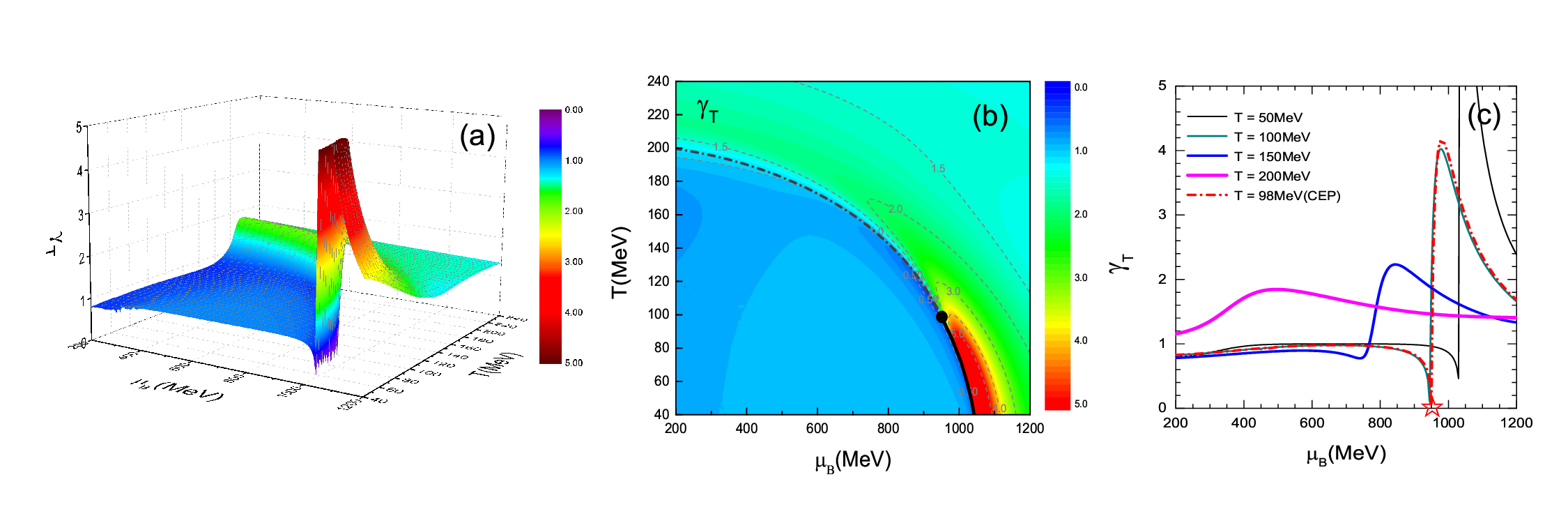}
\caption{The 3D plot (a) and contour map (b) of polytropic index $\gamma_T$ in the $\mu_B-T$ plane based on a 3-flavor PNJL model, as well as the polytropic index $\gamma_T$ as functions of $\mu_B$ (c) at $T = 50$, 100, 150, 200 MeV. In (b), the black dash-dotted and solid lines are respectively for the chiral crossover and first-order phase transition, the black dot connecting the chiral crossover and first-order phase transition represents the critical endpoint (CEP), and the gray dashed lines correspond to the contour of the squared speed of sound. The red star in (c) marks the baryon chemical potential of the CEP.} \label{fig3}
\end{figure*}

\begin{figure*}[tbh]
\includegraphics[scale=0.45]{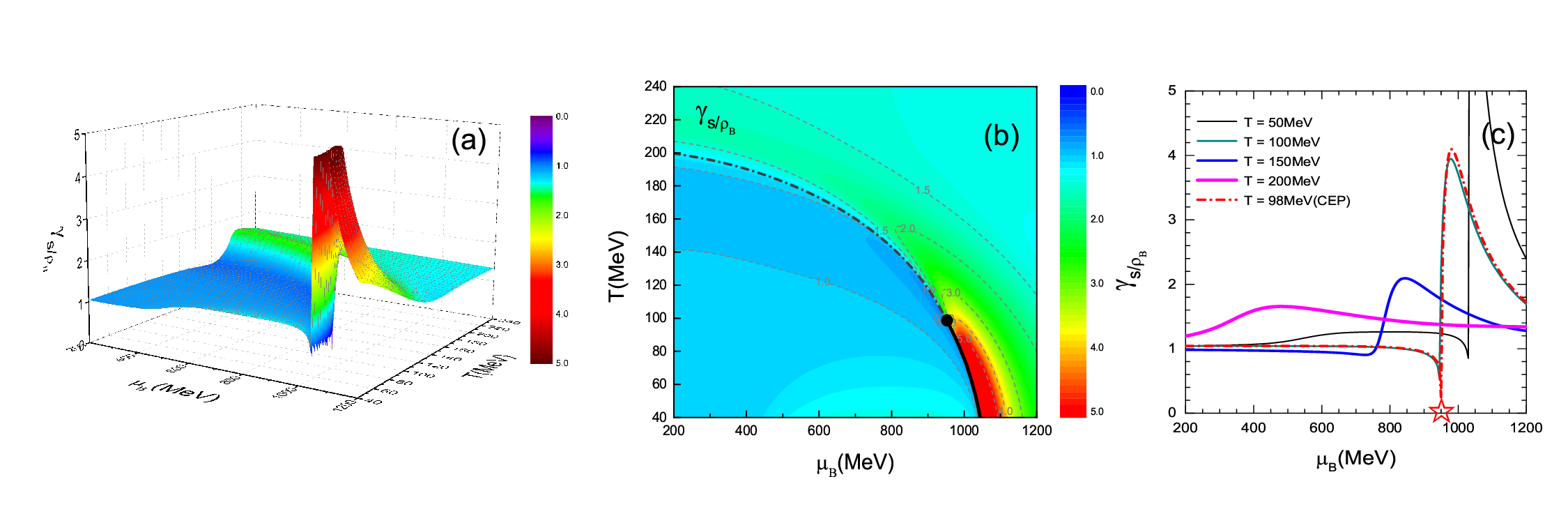}
\caption{The same as in Fig.~\ref{fig3}, but for the polytropic index $\gamma_{s/\rho_B}$. } \label{fig4}
\end{figure*}

We first discuss the squared speed of sound in the $\mu_B-T$ plane based on the 3-flavor PNJL model. As shown in Fig.~\ref{fig1}, we present the 3D plot and contour map of the squared speed of sound $c_T^2$ in panels (a) and (b). The black dash-dotted and solid lines are respectively for the chiral crossover and first-order phase transition, and the black dot connecting the chiral crossover and first-order phase transition represents the CEP. The CEP with $\mu_B=950$ MeV and $T=98$ MeV is found using the present PNJL model and it can vary depending on the parameter space or types of the models. Predictions of CEP from other effective methods can be found in Refs.~\cite{Ste04} and references therein. One can see that the value of $c_T^2$ in the chiral breaking region is mostly less than 0.1. After passing through the chiral phase transition boundary, $c_T^2$ rapidly increases with the restoration of chiral symmetry and gradually approaches the value $c_s^2=1/3$ of conformal limit at the high temperature and baryon chemical potential. It is an expected behavior since the quark condensate rapidly approaches vanishing during chiral restoration, and thus the quark matter at high temperatures and densities within the PNJL model can be considered as non-interacting conformal matter. Based on different effective models and parameter spaces, the similar results for contour map of the square of the adiabatic sound speed can also be found in Ref.~\cite{He22,He23}, where further discussions extend to results under different invariant variables. Different from quark matter, most hadronic models, e.g., an improved isospin- and momentum-dependent interaction (ImMDI) model, indicate that the nucleon interactions are proportional to the baryon density~\cite{Xu15,Xu19}, leading to a steady increase in the speed of sound until the speed of light limit at high baryon density. If the phase transition from hadronic matter to quark matter has taken place, the value of speed of sound could have dramatic changes at a certain density. It is interesting that recent hybrid star research suggests that a sudden downward step change of the speed of sound occurs in the hadron-quark phase transition, and it is restored with the decrease of nucleon and lepton degrees of freedom in the high density quark phase~\cite{Liu237,Liu238}. This dramatic change in the speed of sound may leave a clear and unique signature in the main frequency of the postmerger gravitational wave (GW) spectrum. Another significant feature in Figs. 1(a) and 1(b) is that a dip structure in $c_T^2$ occurs at the low chemical potential side of the chiral phase transition. Especially around the first-order phase transition, the speed of sound rapidly decreases, and a global minimum value appears at the CEP.  At low temperature and high chemical potential, there is a region where the value of $c_T^2$ is relatively large, which is associated with the quarkyonic phase. This can be understood as a result of the fact that the chiral symmetry of light quarks is restored but the strange quark is still in chiral breaking. With the increase of chemical potential, the value of the speed of sound around the first-order phase transition of the strange quark decreases again, followed by another increase in the speed of sound at the higher baryon chemical potential, and eventually approaching the conformal limit after the chiral restoration of strange quark.

To better characterize the variation of the speed of sound at the first-order phase transition, we plot in Fig.~\ref{fig1} (c) the squared speed of sound $c_T^2$ as functions of $\mu_B$ at a series of temperatures. We can see the very different behavior of the speed of sound depending on whether a chiral phase transition occurs, and whether the chiral phase transition occurs through a smooth crossover or though a first-order phase transition. For the curve with $T = 200$ MeV above the chiral phase transition boundary, the speed of sound is monotonically increasing with the chemical potential. For the curve with $T=150$ MeV, the speed of sound displays a spinodal behavior with a small dip around the smooth crossover. For the curve with $T<100$ MeV, we can see a rapid dive and then a steep rise in $c_T^2$ caused by the first-order phase transition. In particular, for the curve with $T=98$ MeV, the point is shown as the red star in panel (c) where the speed of sound in full phase diagram is a global minimum and approaches to almost zero at the chemical potential $\mu_{B} =950$ MeV. 

Similar to Fig.~\ref{fig1}, we also present the results for adiabatic speed of sound $c_{s/\rho_B}^2$ in Fig.~\ref{fig2}. In comparison with $c_T$, we can find that these two types of speed of sound only show some differences at low temperatures and low chemical potentials. In Ref.~\cite{Sor21}, the authors utilize the formulas (2) and (3) to conduct an intuitive analysis of the two types of speed of sound, and their analyses indicate that the values of $c_{s/\rho_B}^2$ and $c_T^2$ largely coincide for $\mu_B/T\gg1$, especially in the limit $T\rightarrow0$, where both expressions for the squared speed of sound can be written as $c^2=1/\mu_B(dP/d\rho_B)_T$. These analyses are consistent with the results of our calculations at low temperature and high chemical potential, as shown in Figs.~\ref{fig1} (b) and~\ref{fig2} (b). For the high-temperature and high-density regime following chiral restoration, where free quarks are formed, both of these speeds of sound approach the conformal limit. As a result, their differences are mainly in the region of low temperature and chemical potential. Meanwhile, our results also point out that although the global minimum values of $c_T^2$ and $c_{s/\rho_B}^2$ occur at the CEP, neither of them are exactly zero, where $c_{s/\rho_B}^2 = 0.01$ is slightly larger than $c_T^2 = 0.002$. The similar non-zero behavior of the adiabatic speed of sound at the CEP also exists in other modeling studies as result of the mean-field approximation~\cite{He22,He23,Mot20}. 

\begin{figure}[tbh]
\includegraphics[scale=0.35]{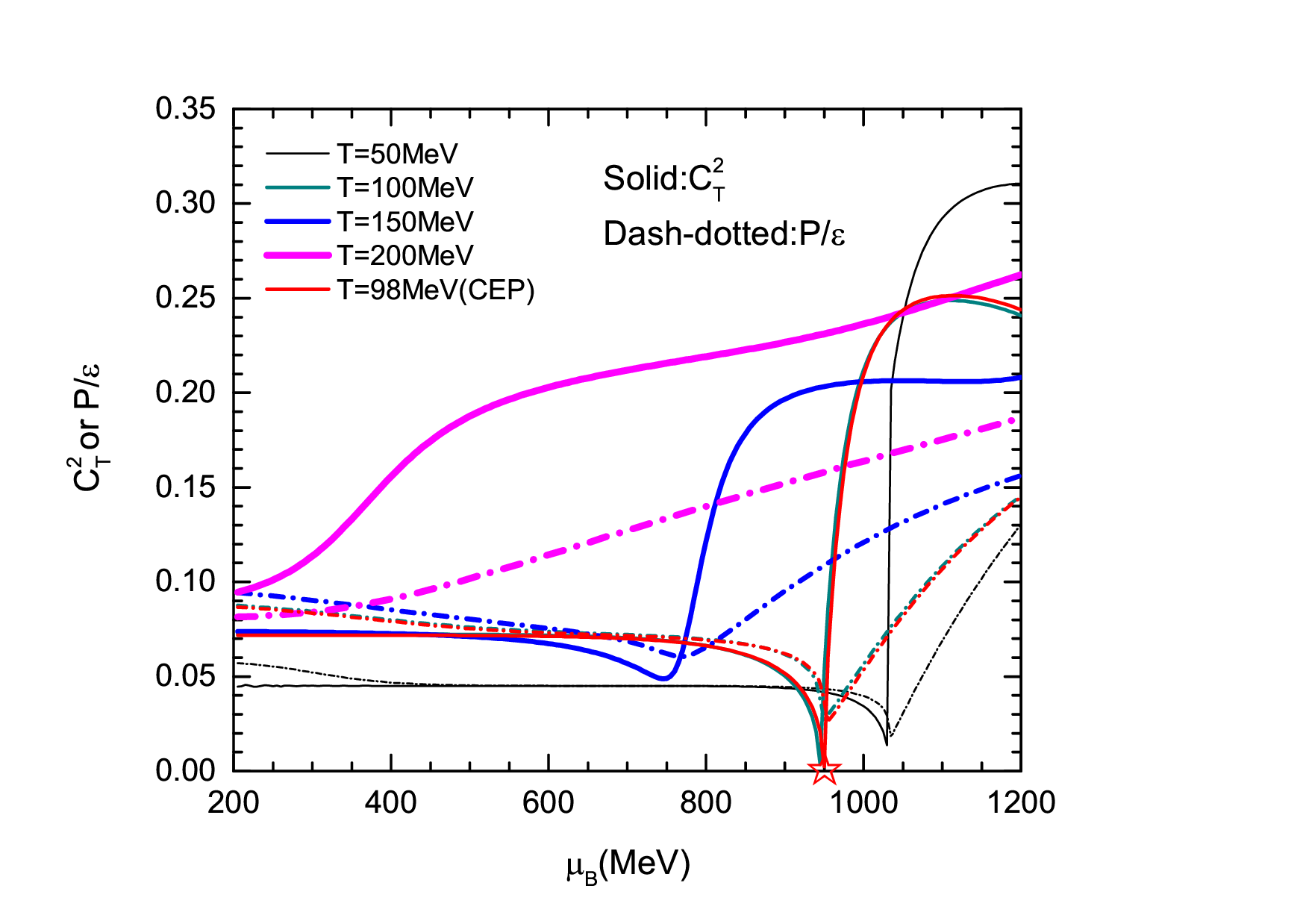}
\caption{The squared speed of sound $c_T^2$ and the factor $P/\varepsilon$ as functions of $\mu_B$ at $T = 50$, 100, 150, 200 MeV. The solid and dash-dotted lines represent the results of $c_T^2$ and $P/\varepsilon$, respectively.} \label{fig5}
\end{figure}

Subsequently, we further investigate the 3D plots and contour maps of the polytropic index $\gamma_T$ and $\gamma_{s/\rho_B}$ as shown in Figs.~\ref{fig3} and~\ref{fig4}. It can be seen that the values of $\gamma_T$ and $\gamma_{s/\rho_B}$ are almost less than 1.5 in the chiral breaking region. However, near the chiral phase transition boundary, the two polytropic indexes quickly drop to a minimum value, then increase rapidly to the maxima with chiral restoration, and subsequently decrease again, eventually reaching the value $\gamma=1$ of conformal limit at high temperature and baryon chemical potential. From the 3D plots in Figs.~\ref{fig3} (a) and~\ref{fig4} (a), we more clearly see the dip and peak structures in the $\gamma_T$ and $\gamma_{s/\rho_B}$ on either side of the phase boundary, especially around first-order phase transition. The polytropic index $\gamma_T$ and $\gamma_{s/\rho_B}$ as functions of $\mu_B$ at various temperatures are illustrated in panel (c) of the Figs.~\ref{fig3} and~\ref{fig4}. For the curve with $T=200$ MeV, different from the monotonic increase of the speed of sound with the chemical potential, both the two polytropic indexes increase to a maximum value and then gradually decrease to the conformal limit. Meanwhile, around the location of the chiral crossover and first-order phase transition, we can also find the nonmonotonic behavior of the $\gamma_T$ and $\gamma_{s/\rho_B}$ with a dip and peak structure. This can be explained by formulas (9) and (10), where the polytropic index is equal to the squared speed of sound divided by the factor $P/\varepsilon$. In Fig~\ref{fig5}, for instance, we plot the squared speed of sound $c_T^2$ and the factor $P/\varepsilon$ as functions of $\mu_B$ at various temperatures. It can be clearly seen that the value of $P/\varepsilon$ basically matches with that of the squared speed of sound for $\mu_B \ll \mu_{B_C}$, where $\mu_{B_C}$ is the baryon chemical potential of the chiral phase transition, and also goes close again at high baryon chemical potential, leading to the fact that the polytropic index is close to 1. Except for these two regions, the squared speed of sound (the derivative of $P$ with respect to $\varepsilon$) does not match $P/\varepsilon$. Around the location of the chiral crossover, as illustrated by the blue lines at $T=150$ MeV in the figure, the local minima of these two quantities are not synchronized, making the minimum and maximum of the polytropic index appear at different chemical potentials, but for the first-order phase transition or the critical point, this desynchronization gradually diminishes, causing the polytropic index to abruptly jump from a near-zero minimum value to its maximum at almost the same chemical potential. Similar to the speed of sound, the two types of polytropic index reach their minimum value at the CEP but are not completely zero, where the value in adiabatic is greater than that in isothermal.

\begin{figure*}[tbh]
\includegraphics[scale=0.48]{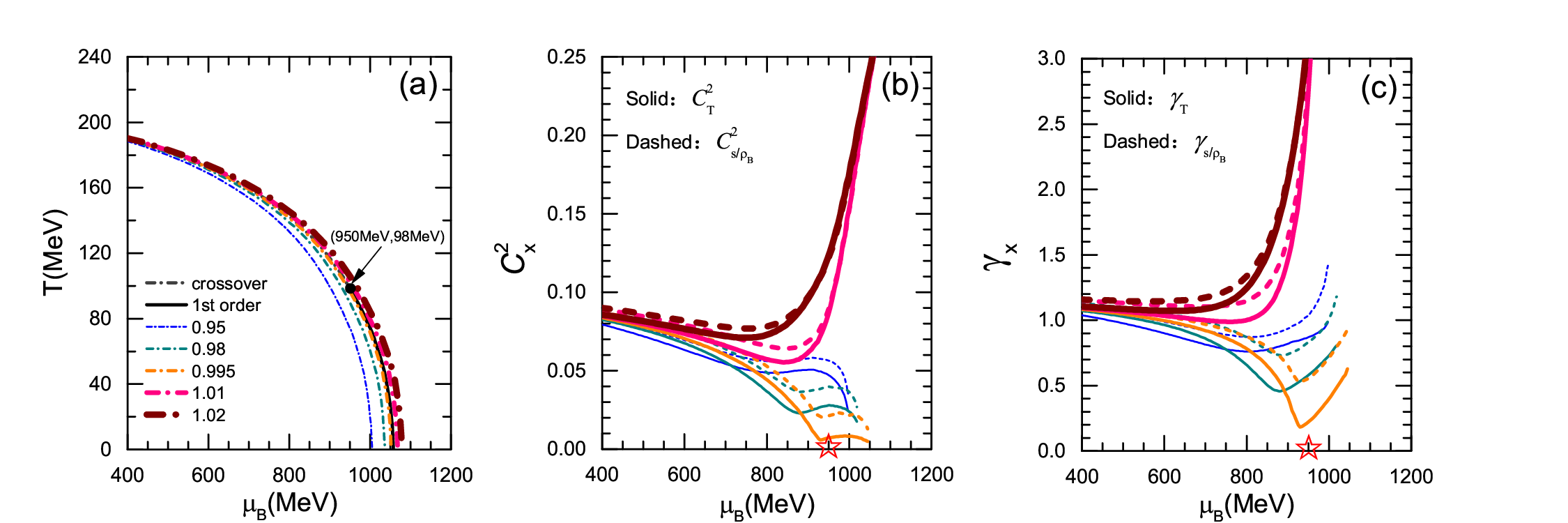}
\caption{The hypothetical chemical freeze-out lines (a) by rescaling $\mu_B$ of the chiral phase transition boundary with factors of 0.95, 0.98, 0.995, 1.01, and 1.02, as well as the squared speed of sound (b) and polytropic index (c) along the hypothetical chemical freeze-out lines. The black dash-dotted and solid lines in (a) are respectively for the crossover and first-order phase transition, but the solid and dashed lines in (b) and (c) represent the results of the speed of sound and polytropic index in isothermal and adiabatic cases, respectively. The black dot in (a) connecting the crossover and first-order phase transition represents the CEP, and the red stars in (b) and (c) mark the baryon chemical potential of the CEP.} \label{fig6}
\end{figure*}

Some studies using the relativistic dissipative hydrodynamics show that the speed of sound as a function of charged particle multiplicity $\langle dN_{ch}/d\eta \rangle$ can be extracted from heavy-ion collision data~\cite{Cam11,Gar20,Sah20}. The recent study in Ref.~\cite{Sor21} also suggests that the net-baryon fluctuations in heavy-ion collisions can be used to estimate the isothermal squared speed of sound  and its logarithmic derivative. In heavy-ion collision experiments, the net-baryon and net-charge fluctuations are measured at the chemical freeze-out. However, the location of the chemical freeze out cannot be well determined at RHIC-BES energies. There are several empirical criteria for the chemical freeze-out, such as fixed energy per particle at about 1 GeV, fixed total density of baryons and antibaryons, fixed entropy density over $T^3$, as well as the percolation model and so on (see reference~\cite{Cle06} and references therein). In order to compare qualitatively the speed of sound and polytropic index from the PNJL model with the (future) experimental results, we obtain the hypothetical chemical freeze-out lines by rescaling $\mu_B$ of the chiral phase transition boundary of light quarks with factors of 0.95, 0.98, 0.995, 1.01, and 1.02 corresponding respectively to the color dash-dotted curves in Fig. 6 (a). The similar assumption of the chemical freeze-out lines was made in reference~\cite{Che16,Che17,Liu21}. We plot in Figs. 6(b) and 6(c) the squared speed of sound and polytropic index along the hypothetical chemical freeze-out lines, where the solid and dashed lines represent the results in adiabatic and isothermal cases, respectively. For the hypothetical freeze-out lines over the chiral phase transition boundary, we can observe a rapid rise in the curves of both the squared speed of sound and polytropic index after a slight softening near the CEP. However, the speed of sound and the polytropic index along the hypothetical freeze-out lines below the chiral phase transition boundary vary considerably: the speed of sound rapidly decreases near the CEP, especially as $c_T^2$ approaches 0 along the curve with the factor of 0.995, followed by a small spinodal behavior and eventually continuing to decrease, while the polytropic index, especially $\gamma_T$, exhibits a more pronounced and nearly closed to zero dip structure as it approaches the CEP. Compared to the speed of sound, the polytropic index could provide a more sensitive probe of the QCD CEP in future experimental exploration.

\section{Summary and Outlook}
\label{SUMMARY}
In conclusion, we have investigated the speed of sound and polytropic index of QCD matter in the full phase diagram based on a 3-flavor Polyakov-looped Nambu-Jona-Lasinio (PNJL) model. In this work, we mainly calculate the speed of sound and polytropic index in the adiabatic and isothermal cases and analyze the changing behavior of these quantities near the CEP. As a result, we can find that a dip structure in both $c_T^2$ and $c_{s/\rho_B}^2$ occurs at the low chemical potential side of the chiral phase transition, and these two types of speed of sound have both reached their global minimum values at the CEP but are not completely zero, where the value in adiabatic is greater than that in isothermal. The polytropic index exhibits similar behavior, but the difference is that it does not increase monotonically along any isothermal curve in the phase diagram. Especially around the location of the chiral crossover and first-order phase transition, we find the dip and peak structures in the $\gamma_T$ and $\gamma_{s/\rho_B}$ on either side of the phase boundary. Along the hypothetical chemical freeze-out lines below the chiral phase transition boundary, the speed of sound rapidly decreases near the CEP, followed by a small spinodal behavior, while the polytropic index, especially $\gamma_T$, exhibits a more pronounced and nearly closed to zero dip structure as it approaches the CEP. Compared to the speed of sound, the polytropic index could provide a more sensitive probe of the QCD CEP in future experiments. However, the polytropic index at present is mainly applied to the study of the massive neutron stars. The approximate rule that $\gamma < 1.75$~\cite{Ann20} or $\gamma\leq 1.6$ and $c_s^2\leq0.7$~\cite{Han23} can be used as a good criterion for separating quark matter from hadronic matter in the massive neutron star core. In a recent study~\cite{Ann23}, the authors introduced a new conformality criterion $d_c <0.2$, defined as being composed of $\gamma$ and $c_s$, and compared it with the speed of sound, polytropic index, and normalized trace anomaly ($\Delta=1/3-c_s^2/\gamma$). The results point out that the conformality criterion $d_c <0.2$ is seen to be considerably more restrictive than the criterion $\gamma <1.75$ for quark matter in compact stars~\cite{Ann23}. Additionally, the authors in Ref.~\cite{Sor21} built a connection between the speed of sound and the cumulants of the baryon number in heavy-ion collisions to detect CEP. Further, it is worth looking forward to some observations related to the speed of sound and polytropic index being found and applied to the search for critical endpoint in heavy-ion collisions and the study of compact star and gravitational wave.

\begin{acknowledgments} 
This work is supported by the National Natural Science Foundation of China under Grants No. 12205158, No. 11975132, and No. 12305148, as well as the Shandong Provincial Natural Science Foundation, China Grants No. ZR2021QA037, No. ZR2022JQ04, No. ZR2019YQ01, and No. ZR2021MA037.
\end{acknowledgments}

\end{document}